%% file: template.tex
\documentclass{article}
\renewcommand\footnotemark{}

\usepackage{arxiv}

\usepackage[utf8]{inputenc} 
\usepackage[T1]{fontenc}    
\usepackage{hyperref}       
\usepackage{url}            
\usepackage{booktabs}       
\usepackage{amsfonts}       
\usepackage{nicefrac}       
\usepackage{microtype}      
\usepackage{caption}
\usepackage{graphicx}
\usepackage{multirow}
\usepackage{amsmath}
\graphicspath{ {./images/} }

\title{Modeling Prejudice and Its Effect on\\ Societal Prosperity}

\author{
 Deep Inder Mohan \textsuperscript{*}\thanks{\textsuperscript{*} Equal contributors.}\\
  IIIT Bangalore\\
  Bangalore, India\\
  \texttt{deepinder.mohan@iiitb.ac.in} \\
   \And
 Arjun Verma \textsuperscript{*}\\
  IIIT Bangalore\\
  Bangalore, India\\
  \texttt{arjun.verma@iiitb.ac.in} \\
  \And
 Shrisha Rao \thanks{  Code repository: \url{https://github.com/deep-guy/prejudice_model}} \\
  IIIT Bangalore\\
  Bangalore, India\\
  \texttt{shrao@ieee.org} \\
}

\begin{document}
\maketitle
\begin{abstract}
 Existing studies on prejudice, which is important in multi-group dynamics in societies, focus on the social-psychological knowledge behind the processes involving prejudice and its propagation. We instead create a multi-agent framework that simulates the propagation of prejudice and measures its tangible impact on the prosperity of individuals as well as of larger social structures, including groups and factions within. Groups in society help us define prejudice, and factions represent smaller tight-knit circles of individuals with similar opinions. We model social interactions using the Continuous Prisoner's Dilemma (CPD) and a type of agent called a prejudiced agent, whose cooperation is affected by a prejudice attribute, updated over time based both on the agent’s own experiences and those of others in its faction.  Our simulations show that modeling prejudice as an exclusively out-group phenomenon generates implicit in-group promotion, which eventually leads to higher relative prosperity of the prejudiced population. This skew in prosperity is shown to be correlated to factors such as size difference between groups and the number of prejudiced agents in a group. Although prejudiced agents achieve higher prosperity within prejudiced societies, their presence degrades the overall prosperity levels of their societies. Our proposed system model can serve as a basis for promoting a deeper understanding of origins, propagation, and ramifications of prejudice through rigorous simulative studies grounded in apt theoretical backgrounds. This can help conduct impactful research on prominent social issues such as racism, religious discrimination, and unfair immigrant treatment. This model can also serve as a foundation to study other socio-psychological phenomena in tandem with prejudice such as the distribution of wealth, social status, and ethnocentrism in a society. 
\end{abstract}


\section{Introduction}
\label{section:intro}

Intergroup discrimination is a feature of most modern societies~\cite{tajfel1970}. People typically seek other people who are similar to themselves~\cite{fiske}, which results in these similar individuals to form \textit{groups}. These groups labeled as in-groups, give rise to an individual's social identity~\cite{tajfel1974}. The formation of such an in-group causes the simultaneous creation of one or more out-groups, which comprises of all agents that do not belong to the in-group. Since in societies, this division into groups most often implies a competitive relationship between the groups, the existence of classification is sufficient for the introduction of intergroup bias~\cite{tajfel1970}. This intergroup bias may be exhibited in the form of in-group favoritism, or out-group prejudice~\cite{messick}. 

Out-group prejudice, or simply \textit{prejudice}, is a phenomenon where a group is perceived as a homogeneous entity~\cite{MASUDA20128}, and a cognitive error is directed towards every one of its members simply because of their membership~\cite{allport}. Prejudice may manifest itself through discriminatory behavior towards members of the out-group by certain individuals~\cite{tajfel1970}, which may have a basis in both the individual's personal experience~\cite{fareri} and its tendency to conform to its in-group~\cite{messick}. Experimental studies have shown that certain properties of groups may be correlated with the effects of prejudice in society, such as their sizes ~\cite{Martinez-Vaquero}~\cite{giles}, and even their sociopolitical ordering based on mainstream culture~\cite{dasgupta_implicit_2004}.  Since the formation of groups and the subsequent birth of prejudice is inevitable in any society, understanding prejudice, its manifestation in social interactions, and its effects on individual and collective prosperity become paramount, especially in consideration of current socio-political contexts. 

Past experiments~\cite{tajfel1970} have concluded that prejudice brings about a disparity in the performance of groups. However, we observe that a question on the effects of prejudice remains unaddressed: does this disparity arise only from the prejudiced agents’ explicit suppression (negative) of the out-group, or does its presence in agents also cause promotion (positive) in their in-group? We address this question through a complex model of prejudiced multi-group societies in which we can test the in-group and out-group effects of prejudice independently. This is achieved by instantiating prejudiced groups in society independent of groups that are subject to this prejudice.

The interactions in our model are governed by the Continuous Prisoner’s Dilemma (CPD)~\cite{verhoeff}. These interaction types have recently been shown to be successful in modeling and understanding other social phenomena such as ethical behavior~\cite{ijcai2020-24} and egocentric bias~\cite{nkishore2019} in individuals. Our agent society contains two social structures, groups and factions. Groups represent subsets of agents in the society that share similar key characteristics. Factions are subsets of agents within the same group that have the ability to influence the actions of the constituent faction members. The level of cooperation of an agent in a CPD interaction is governed by its opinion of the other agent. An agent forms its opinion based on its past experience and the opinions of its faction members. The level of influence that an agent's faction has on its opinion is controlled by an attribute named \textit{faction alignment}.

Recent work on understanding the mechanisms of prejudice~\cite{Whitaker2018} models its origins and the societal conditions that lead to its perpetuation and mitigation. Their work assumes that an agent has the same prejudice value against all other out-groups in the society which may not be true. Also, agents in their model take individual actions based on global reputations of other agents, which may not accurately represent the behavior of individuals, especially in prejudiced societies where agents may act in lockstep with only their in-group~\cite{messick}. 

To better understand the effects of prejudice on prosperity, we introduce an agent type called the \textit{prejudiced agent}. We model prejudice as an individual characteristic of this agent where the level of prejudice that any agent can have against a group lies in the range [0, 1]. Each such agent can have different prejudice values against different groups in the society. This prejudice value serves as a discount in the cooperation level of these agents for interactions with the out-group, and can dynamically change based on the payoffs that it receives from these interactions. The change in its prejudice value further propagates a change in its faction alignment based on the difference between its own prejudice value and its faction’s prejudice value against a particular group.

We generate our results after running experiments with varying configurations of our agent society. Result \textbf{R2} shows that even when modeling prejudice as having an explicit effect only on inter-group interactions, there is still an emergence of implicit in-group promotion. We conclude that the existence of out-group suppression and in-group promotion cannot be isolated, although, in our model, the former has a more significant impact on the creation of disparity within society than the latter.

We demonstrate through result \textbf{R1} that groups with a higher concentration of prejudiced agents in them demonstrate higher prosperity than their less prejudiced counterparts. Our modeling of imbalanced societies also demonstrates a correlation between skew in sizes and skew in prosperity levels of competing groups, which is further amplified by the existence of prejudice in the majority group. Furthermore, result \textbf{R4} addresses the effects of prejudice in skewed societies, and \textbf{R5} summarises the impact of prejudice on the prosperity of a society as a whole.

We also present results for societies containing agents whose prejudice may be directed towards the in-group, rather than the out-group. These type of agents may arise in socially disadvantaged groups~\cite{spicer} We model these agents as \textit{renegades}, who discriminate against members of their own group on the basis of prejudice. Existing research on renegade agents~\cite{dasgupta_implicit_2004} hypothesizes that these agents may have lower levels of prosperity within society, and may even reduce the prosperity levels of others within their group. Our result \textbf{R3} validates these hypotheses and further presents findings into trends followed by such agents’ prejudice levels over time.

\section{System Model}
\label{section:fw}

We model a system to represent an agent society in which the constituent agents participate in iterated interactions. The set of all agents in the society is denoted by $\mathbb{S}$. At each model step, any two agents are picked at random from $\mathbb{S}$ and paired together to undergo an interaction. 

\subsection{Agent Experience and Opinion}

Each agent has a notion of memory ingrained in them. The memory serves the purpose of storing the most recent experiences of an agent with all other agents that it has interacted with until that instance. This is achieved using a matrix $\mathbb{E}$, termed as \textit{agent experience}. The size of $\mathbb{E}$ can be of the order $\mathcal{O}(n \cdot \omega)$, where \textit{n} is the total number of agents in the society and \textit{$\omega$} represents the maximum number of recent interactions that an agent can recall with another agent.

\textit{Agent experience} $(\mathbb{E})$ is utilised by an agent to form an \textit{opinion} about the other agent in a particular interaction~\cite{nkishore2019}. The interactions in the model are based on Continuous Prisoner's Dilemma (CPD) which is elaborated upon in Section \ref{subsec:cpd}. We now consider two randomly picked agents, $\mathcal{A}{}_0$ and $\mathcal{A}{}_1$ in an arbitrary interaction. We use $C_{\mathcal{A}{}_0}(\mathcal{A}{}_1,t)$ to denote $\mathcal{A}{}_0$'s \textit{cooperation level} with $\mathcal{A}{}_1$ at interaction $t$ between them. The \textit{cooperation level} is a value that lies in the range of 0 to 1 as seen in standard CPD settings. The \textit{opinion} of $\mathcal{A}{}_0$ about $\mathcal{A}{}_1$ at interaction \textit{t} between the two is denoted by $\eta_\mathcal{A}{}_0(\mathcal{A}{}_1, t)$. This value is computed as,

\begin{equation} \label{eq : 1}
    \eta_{\mathcal{A}{}_0}(\mathcal{A}{}_1,t) = \frac{\sum_{i=1}^{\omega} C_{\mathcal{A}{}_0}(\mathcal{A}{}_1, t-i)}{\omega}
\end{equation}

We also define social structures: \textit{groups} and \textit{factions} in our model. Each agent belongs to a particular \textit{group} in the system and a \textit{faction} within the \textit{group}. A \textit{faction} is a small subset of agents within the  \textit{group} that represents a circle of close-knit agents. These are discussed in detail in Section \ref{subsec:socialstructures}. If an agent belongs to a \textit{faction}, the final cooperation level $C_{\mathcal{A}{}_0}(\mathcal{A}{}_1,t)$ is then an aggregation of $\mathcal{A}{}_0$'s past experience with $\mathcal{A}{}_1$, ($\eta_{\mathcal{A}{}_0}(\mathcal{A}{}_1, t)$) and the opinion of $\mathcal{A}{}_0$'s faction of $\mathcal{A}{}_1$. 

\subsection{Continuous Prisoner’s Dilemma}
\label{subsec:cpd}
 
In a standard Prisoner's Dilemma (PD) game setting, the agents have a choice to either defect or cooperate with the other agent which in turn affects their payoffs. Such extreme behaviors of only cooperation or defection however cannot model complex real-word scenarios. This can be achieved by using an alternate game setting: the Continuous Prisoner's Dilemma (CPD)~\cite{verhoeff}, which allows interacting agents to choose a level of cooperation in the range of 0 to 1, where 0 indicates complete defection and 1 indicates complete cooperation. Consider an agent $\mathcal{A}{}_0$ interacting with another agent $\mathcal{A}{}_1$. Their cooperation levels are denoted by $c_0$ and $c_1$ respectively. $\mathcal{A}{}_0$'s payoff, $r_{\mathcal{A}{}_0}(c_0,c_1)$, is now calculated as a linear interpolation of the discrete game payoffs
    \begin{equation} \label{eq : 2}
            r_{\mathcal{A}{}_0}(c_0,c_1) = {c_0}{c_1}C + {c_0}\bar{c_1}S + \bar{{c_0}}{c_1}T + \bar{{c_0}}\bar{c_1}D
    \end{equation}
    where $\bar{c_0} = (1-c_0)$, $\bar{c_1} = (1-c_1)$, and $C, T, D, S$ are the discrete payoffs in  standard PD as shown below. 
    \input{tables/cpd}
    
The conditions to be followed while choosing the values of these variables are : $2C > T+S$ and $T > C > D > S$. Following the work done in~\cite{Axelrod1390}~\cite{nkishore2019} we choose the set of values to be : $(C = 3, T = 5, D = 1, S = 0)$   

\subsection{Social Structures} 
\label{subsec:socialstructures}
    The fundamental social structure in our model is a \textit{group}, which represents a subset of agents in the society who share similar key values, and group membership is a part of an agent's social identity. All agents in $\mathbb{S}$ belong to a particular \textit{group}. The \textit{groups} in a society are mutually exclusive and exhaustive, i.e,  $G_1$ $\cup$ $G_2$ $\cup$ ... $\cup$ $G_g$ $=$ $\mathbb{S}$ and $G_1$ $\cap$ $G_2$ $\cap$ ... $\cap$ $G_g$ $=$ $\phi$, where g is the total number of groups in society.
    
    Additionally, every agent in a group also belongs to a particular \textit{faction} within it. Since agents in a group may have a direct relationship with only a subset of their group~\cite{allport}, we create \textit{factions} to model these subsets, where opinions of agents are influenced by others in their faction. The set of factions, \{$\mathcal{F}_1$, $\mathcal{F}_2$, ..., $\mathcal{F}_f$\}, is mutually exclusive and exhaustive, where \textit{f} represents the total number of factions within a \textit{group}. Consider an agent $\mathcal{A}{}_0$ interacting with another agent $\mathcal{A}{}_1$. While calculating its level of cooperation with $\mathcal{A}{}_1$, $\mathcal{A}{}_0$ takes into consideration not only its own \textit{opinion} (as computed in \eqref{eq : 1}) of $\mathcal{A}{}_1$ but also an aggregate of the opinions of its faction members. Let $\mathcal{F}$ represent the faction to which an agent $\mathcal{A}{}_0$ belongs. The faction's \textit{aggregated opinion} of $\mathcal{A}{}_1$ at interaction $t$, denoted by $\mathcal{O}_\mathcal{F}(\mathcal{A}{}_1, t)$, is thus computed as,
    
    \begin{equation} \label{eq : 3}
        \mathcal{O}_\mathcal{F}(\mathcal{A}{}_1, t) = \frac{\sum_{\forall \mathcal{A} \in \mathcal{F}} \eta_\mathcal{A}(\mathcal{A}{}_1,t)}{|\mathcal{F}|}
    \end{equation}
    
    The level of influence that an agent's faction can have over its \textit{opinion} is modeled by a parameter $f$ called the \textit{faction alignment}. It denotes the weightage that $\mathcal{A}{}_0$ gives to its faction's opinion. Its value lies between 0 and 1, with $f=0$ representing that $\mathcal{F}$ has no influence on $\mathcal{A}{}_0$'s opinion of $\mathcal{A}{}_1$, and $f=1$ representing that $\mathcal{A}{}_0$ has complete faith in $\mathcal{F}$ and decides its level of cooperation entirely based on its faction's opinion.
    
    \subsection{Prejudiced Agent}
\label{subsec:prej}
In this section, we introduce the prejudiced agent type. For an agent $\mathcal{A}$, prejudice is modeled as a vector named \textit{prejudice vector}: [$p_1$, $p_2$, ..., $p_g$], where \textit{g} is the total number of \textit{groups} in the system and $p_i$ is a value between 0 to 1 representing the level of prejudice that $\mathcal{A}$ has against the members of \textit{group} $G_i$. The value of $p_i$ is updated after every interaction $\mathcal{A}$ has with an agent belonging to $G_i$, where the outcome of such interactions could either reinforce or quell $\mathcal{A}$'s prejudice.

Based on its prejudice value and the prejudice values of the members of its faction, a prejudiced agent can also update its \textit{faction alignment} $f$. \textit{Faction alignment} updates are modeled to allow the prejudiced agents to dynamically change the effect of their faction's influence over it. A prejudiced agent's faction would have more influence over it if the agent's prejudice value is close to the average prejudice value of the rest of its faction members since this implies that the agent and its faction share a similar world view. These updates are discussed further in detail in Section \ref{subsec:interactions}. Since an unprejudiced agent has no notion of prejudice in it, these updates are not required for such agents. The $p_i$'s of a prejudiced agent's \textit{prejudice vector} are instantiated at random using a Gaussian Distribution defined by $\mu$ = 0.5 and $\sigma$ = 0.2. The use of a Gaussian Distribution to instantiate the prejudice vector is in line with previous work on other socio-psychological factors ~\cite{nkishore2019} ~\cite{Lieder}. We now consider a prejudiced agent $\mathcal{A}{}_0$ belonging to \textit{group} $G_0$ having its interaction $t$ with another agent $\mathcal{A}{}_1$ of a different \textit{group} $G_1$. The prejudice attribute of $\mathcal{A}{}_0$ influences its \textit{opinion} (as computed in \eqref{eq : 1}) about $\mathcal{A}{}_1$ and this new \textit{prejudiced opinion}, $\Psi_{\mathcal{A}{}_0}(\mathcal{A}{}_1,t)$ is thus computed as,

\begin{equation} \label{eq : 4}
        \Psi_{\mathcal{A}{}_0}(\mathcal{A}{}_1,t) = (1 - p_1) \cdot \eta_{\mathcal{A}{}_0}(\mathcal{A}{}_1,t)
\end{equation}

In a faction comprising of prejudiced agents, the collective opinion is shaped by the prejudice of agents contributing to its formation~\cite{brewer1998intergroup}. Therefore, the aggregated opinion is no longer computed using \eqref{eq : 3}. Instead, we use \eqref{eq : 4} to calculate the new \textit{prejudiced aggregated opinion}, $\mathcal{P}_\mathcal{F}(\mathcal{A}{}_1, t)$ as,

\begin{equation} \label{eq : 5}
        \mathcal{P}_\mathcal{F}(\mathcal{A}{}_1, t) = \frac{\sum_{\forall \mathcal{A} \in \mathcal{F}} \Psi_\mathcal{A}(\mathcal{A}{}_1,t)}{|\mathcal{F}|}
\end{equation}


\subsection{Agent Interactions}
\label{subsec:interactions}

The \textit{cooperation level} computed by an agent undergoing a CPD interaction changes depending on whether the agent is prejudiced, or unprejudiced. We describe the method for the computation of this value in both these agent types. 

\subsubsection{Unprejudiced Agent Interaction}
\label{subsec:unprej}

The \textit{cooperation level} for an unprejudiced agent is computed by combining its \textit{opinion}, as derived from \eqref{eq : 1} and its faction's \textit{aggregated opinion}, as derived from \eqref{eq : 3}. Consider an unprejudiced agent $\mathcal{A}{}_0$ interacting with another agent $\mathcal{A}{}_1$. Let the \textit{faction alignment} (defined earlier in Section \ref{subsec:cpd}) of $\mathcal{A}{}_0$ be $f$. $f$ governs how much influence $\mathcal{A}{}_0$'s faction's \textit{aggregated opinion} has on its \textit{cooperation level}. The \textit{cooperation level} for $\mathcal{A}{}_0$ at interaction \textit{t} with $\mathcal{A}{}_1$ is then computed as,

\begin{equation} \label{eq : 6}
        \mathcal{C}_{\mathcal{A}{}_0}(\mathcal{A}{}_1, t) = f \cdot \mathcal{O}_\mathcal{F}(\mathcal{A}{}_1, t) + (1-f) \cdot \eta_{\mathcal{A}{}_0}(\mathcal{A}{}_1,t)
\end{equation}

The payoff obtained by $\mathcal{A}{}_0$ can be calculated by substituting its \textit{cooperation level} in \eqref{eq : 2} and this payoff can then be cumulated over every interaction to serve as the agent's measure of \textit{prosperity}.

\subsubsection{Prejudiced Agent Interaction}

The \textit{cooperation level} for a prejudiced agent is computed in a similar but slightly different manner than the unprejudiced agent. Consider a prejudiced agent $\mathcal{A}{}_0$ belonging to \textit{group} $G_0$ interacting with another agent $\mathcal{A}{}_1$ belonging to \textit{group} $G_1$. The prejudice value of $\mathcal{A}{}_0$ against members of $G_1$ is denoted by $p_1$. Let the \textit{faction alignment} of $\mathcal{A}{}_0$ be $f$. The \textit{cooperation level} for $\mathcal{A}{}_0$ at interaction \textit{t} with $\mathcal{A}{}_1$ is then computed using \eqref{eq : 4} and \eqref{eq : 5} as,

\begin{equation} \label{eq : 7}
        \mathcal{C}_{\mathcal{A}{}_0}(\mathcal{A}{}_1, t) = f \cdot \mathcal{P}_\mathcal{F}(\mathcal{A}{}_1, t) + (1-f) \cdot \psi_{\mathcal{A}{}_0}(\mathcal{A}{}_1,t)
\end{equation}

\noindent This is again substituted in \eqref{eq : 2} to calculate the payoff received which is then cumulated to get the agent's measure of \textit{prosperity}. The major distinction in prejudiced and unprejudiced interactions comes after the calculation of \textit{cooperation level}. As discussed briefly in Section \ref{subsec:prej}, a prejudiced agent now updates both, its prejudice value $p_i$ as well as its \textit{faction alignment} $f$ based on the payoff received from the CPD interaction.

If the payoff received by $\mathcal{A}{}_0$ goes above a threshold value $\alpha_1$, its prejudice value $p_i$ is incremented by a value $\theta$ as its prejudice is reinforced due to the rewarding payoff. Similarly, if it goes below a threshold value $\alpha_2$, its prejudice value is decremented by the same $\theta$ as in this case the agent acted upon its prejudice but received a lower payoff which decreases its belief in its prejudice.

\begin{equation} \label{eq : 8}
            p_i := p_i   
            \begin{cases}
                + \theta & \text{if \textit{Payoff} $> \alpha_1 $ } \\
                - \theta & \text{if \textit{Payoff} $< \alpha_2 $ }
            \end{cases}
\end{equation}

\noindent Varying fractions of the maximum payoff (= 5) that an agent could obtain from a standard CPD interaction (obtained when one agent fully cooperates and the other fully defects) were tested as values for the thresholds, $\alpha_1$ and $\alpha_2$, while keeping $\alpha_1$ $ > $ $\alpha_2$. The final values for $\alpha_1, \alpha_2 \text{ and } \theta$ are given in Table \ref{table:model}.

After updating its prejudice, $\mathcal{A}{}_0$ now updates its \textit{faction alignment} $f$. This is done by first computing a new quantity \textit{$Prej_F$}, termed as \textit{faction prejudice}. $Prej_F$ is computed as the average of $p_i$'s of all the members of the faction. In this particular interaction, the value of $Prej_F \text{ is } \frac{\sum_{\forall \mathcal{A} \in \mathcal{F}}p_1}{|\mathcal{F}|}$. If the difference between the agent's own prejudice and the \textit{faction prejudice} is less, more specifically, lower than a threshold $\beta_1$, the agent reinforces its faith in its faction by increasing its \textit{faction alignment} $f$ by a value $\tau$. Similarly, if the difference is greater than a threshold $\beta_2$, it decreases its \textit{faction alignment} by the same value $\tau$. In this particular interaction, the update is shown below as,

\begin{equation} \label{eq : 9}
            f := f   
            \begin{cases}
                + \tau & \text{if $|Prej_F - p_1| < \beta_1$ } \\
                - \tau & \text{if $|Prej_F - p_1| > \beta_2$}
            \end{cases}
\end{equation}

\input{tables/table_1}

As the value for an agent's prejudice updates over interactions, the thresholds $\beta_1$ and $\beta_2$ cannot be kept constant. They are computed dynamically using an agent's current \textit{FactionAlignment} $f$. By tracking the value of  $|Prej_F - p_i|$ in multiple simulations, we find it to be in the range [0.01, 0.2]. This range is then further split to determine the ranges for $\beta_1$ and $\beta_2$, where $\beta_1 < \beta_2$. These thresholds for an agent $\mathcal{A}$ are then recalculated before every interaction by linearly scaling $\mathcal{A}$'s current value of $(1-f)$ to the predetermined range of each threshold. The value of $\tau$ and the ranges for $\beta_1$ and $\beta_2$ after multiple runs of the simulation are shown in Table \ref{table:model}.

\input{tables/table_2}
\section{Experiments and Results}

Each instantiation of the model containing agents belonging to \textit{groups} and \textit{factions} is termed as a \textit{society}. 
Every \textit{society} contains 1000 agents, which may either be prejudiced or unprejudiced, as described in Section \ref{section:fw}. Societies have exactly 50 \textit{factions}, each containing 20 agents of the same type. Each society may contain two or more \textit{groups}, and all agents and the corresponding factions they belong to are distributed uniformly across these groups unless specified otherwise. Prejudice orientations are taken to be a group trait, therefore every prejudiced agent in a group is prejudiced towards the same set of out-groups.

Experiments are run for $10^5$ iterations, where every iteration is a CPD interaction (Section \ref{subsec:cpd}) between two agents chosen randomly from the agent pool. Each experiment is repeated 10 times, and the results generated are the average over these runs. Experiments are run with the values of model parameters as specified in Table \ref{table:model}.

In our experiments, we calculate the cumulative payoff (labeled \textit{prosperity}) of each agent and use it to generate results on average \textit{prosperity} values of different groups/agent types.
We also track prejudice and payoff values of our agent body after every iteration, and present results on the change of these quantities over time.

\subsection{Effect of Prejudice on \textit{Prosperity}}
\label{subsec:3.1}
    \subsubsection{Concentration of Prejudice}
    
    \begin{figure}[h]
        \centering
        \includegraphics[width=0.45\textwidth]{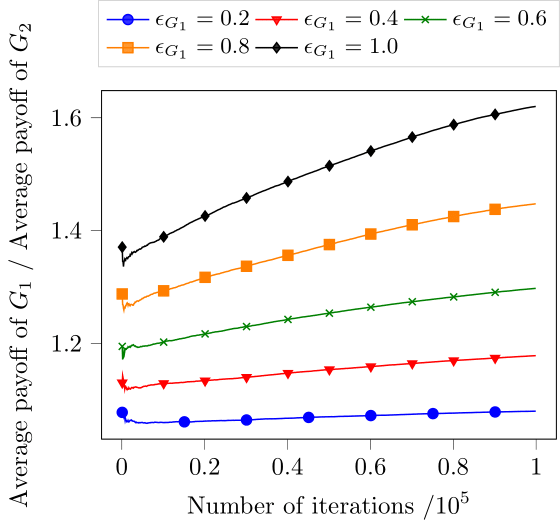}
        \caption{Payoff over time for different concentrations of prejudiced agents}
        \label{fig:4.1.1}
    \end{figure}
    
    \label{subsub:3.1.1}
    We analyze the effect of the fraction of agents which are prejudiced in a group $G$, denoted as $\epsilon_G$ on the relative performance of that group over time. $\epsilon_G$ can be interpreted as a measure of the level of prejudice of $G$.
     We initialize a model with two groups, $G_1$ and $G_2$, where all agents in $G_2$ are unprejudiced ($\epsilon_{G_2} = 0$), and we test for 5 fractions of  prejudiced agents in $G_1$, $\epsilon_{G_1} \in \{0.2, 0.4, 0.6, 0.8, 1.0\}$. The ratio of the average payoff of agents in $G_1$ and that of agents in $G_2$ at every model step is plotted in Fig \ref{fig:4.1.1} for all 5 values of $\epsilon_{G_1}$.
    
    As the plots indicate, the payoff ratio is directly correlated to the value of $\epsilon_{G_1}$. This brings us to our first result:
    
    \begin{center}
        \fbox{%
        \parbox{0.8\columnwidth}{%
            \textbf{R1.} \emph{The relative average prosperity of a group within a society increases as the group gets more prejudiced.} 
        }%
        }
    \end{center}

    This result is in agreement with existing studies on prejudiced behavior, and also provides an incentive for why prejudice may exist in societies in the first place.
    
    \subsubsection{In-Group Effect vs. Out-Group Effect}
    
    \begin{figure}[h]
      \centering
      \begin{minipage}[t]{.45\textwidth}
      \centering
      \vspace{-12.7em}
        \include{tables/table_3}
        \captionof{table}{Prejudice orientation for results in Fig \ref{fig:bias}}
        \label{table:tab1}
      \end{minipage} \qquad
      \begin{minipage}[t]{.45\textwidth}
        \centering
        \includegraphics[width=0.85\textwidth]{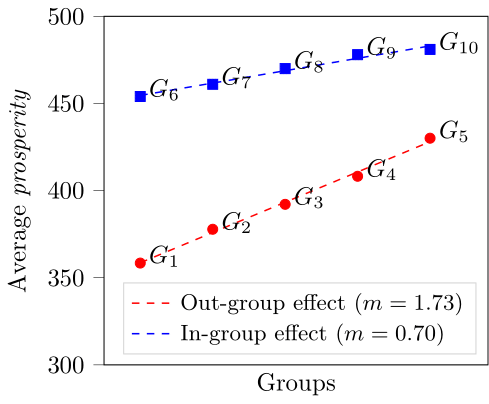}
        \captionof{figure}{Comparing in-group and out-group effects of prejudice ($m$ is the slope)}
        \label{fig:bias}
      \end{minipage}%
    \end{figure}
    
    Does the disparity presented in Fig \ref{fig:4.1.1} arise only from the prejudiced agents’ explicit (negative) suppression of the out-group, or does its presence in agents also bring about implicit (positive) promotion in their in-group?
    
    In order to address this question, we run an experiment as follows: we initialize our society with 10 groups, $\{G_1, ..., G_{10}\}$, where every group has 100 agents. Groups $G_1, ..., G_5$ are identical, containing all unprejudiced agents, whereas groups $G_6, ..., G_{10}$ have all prejudiced agents.
    Prejudice orientations of these agents are initialized as in Fig \ref{table:tab1}, where `Groups Prejudiced against' lists all the groups that the given group is prejudiced against. Therefore, $G_{10}$ is prejudiced against 5 groups, and conversely, $G_1$ has 5 groups prejudiced against it.
    We use this initialization of our prejudice orientation for a specific reason: this orientation creates a trend for decreasing out-group effects $(G_1 > ... > G_5)$ and increasing in-group effects $(G_6 < ... < G_{10})$ of prejudice simultaneously within the same society but in independent sets of groups.

    Fig. \ref{fig:bias} present the results for each of these effects, and we observe a positive slope for both effects, indicating that they exist in tandem. However, we note that the increase in \textit{prosperity} of unprejudiced groups due to decreasing prejudice against them is more severe $(m=1.7)$ than the increase in \textit{prosperity} of the prejudiced groups due to their increasing level of prejudice $(m=0.7)$. This brings us to the following: 
    
    \begin{center}
       \fbox{%
        \parbox{0.8\columnwidth}{%
            \textbf{R2.} \emph{In-group promotion due to the existence of prejudice emerges implicitly in societies even when prejudice is modeled explicitly as an out-group phenomenon, although the severity of this effect on the creation of disparity between groups is lower.}
        }%
        }
    \end{center}
    
\subsection{Majorities and Minorities}
\label{subsec:3.2}

    \begin{figure}[h]
      \centering
      \begin{minipage}[c]{.45\textwidth}
        \centering
        \includegraphics[width=\textwidth]{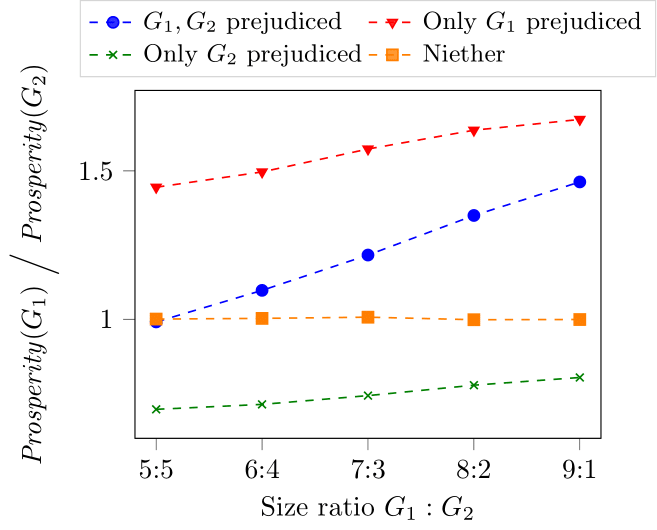}
        \captionof{figure}{\textit{Prosperity} ratios for multiple configurations of skewed societies}
        \label{fig:4.1.3}
      \end{minipage}\qquad
      \begin{minipage}[c]{.45\textwidth}
        \centering
        \vspace{1.5em}
        \includegraphics[width=0.80\textwidth]{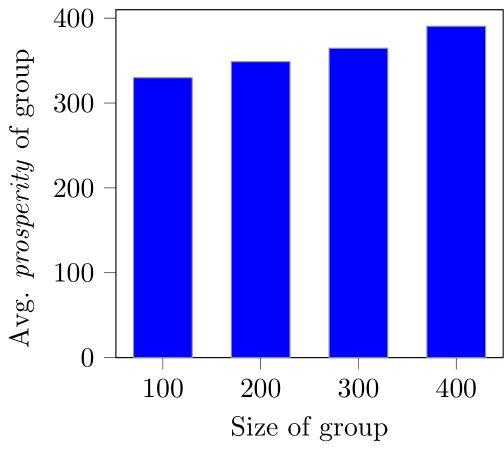}
        \captionof{figure}{\textit{Prosperity} variations in a skewed society}
        \label{fig:4.1.4}
      \end{minipage}
    \end{figure}

    We test the effect that the relative size of a group has on its relative \textit{prosperity}, and analyze the impact of the existence of majorities and minorities in society. For this, we conduct a series of experiments for societies having two groups $G_1$ and $G_2$, and study changes in the ratio of the average \textit{prosperity} of these groups with changes in skew between their sizes. We emulate societies with four prejudice configurations of our groups: Both $G_1$ and $G_2$ prejudiced; only $G_1$ prejudiced; only $G_2$ prejudiced; and both $G_1$ and $G_2$ unprejudiced. For each of these configurations, we run simulations for 5 different splits of agents between these groups: 500-500; 600-400; 700-300; 800-200; 900-100.
    
    The values of the ratio of the average \textit{prosperity} of groups $G_1$ and $G_2$ from these experiments are plotted in Fig \ref{fig:4.1.3}. The figure shows that while skews in size have no impact on \textit{prosperity} when neither group is prejudiced (orange), we can observe a significant increase in \textit{prosperity} of the agents in the majority group over those in the minority group in the configurations where the majority is prejudiced (red, blue). A relative increase in prosperity of the majority can also be observed in the case when only the minority is prejudiced (green), although this is much more subtle. However, even with a 9:1 skew in group size in favor of the unprejudiced group, the prejudiced group still maintains a higher prosperity level. This brings us to the following result: 
    
    \begin{center}
        
    \fbox{%
    \parbox{0.8\columnwidth}{%
        \textbf{R3.} \emph{Skew in group size in prejudiced societies results in a relative increase in prosperity of the majority. This skew creates a disparity in favor of the majority group when it is prejudiced, but cannot overcome the existing disparity from the prejudice of the minority group when it is unprejudiced.}
    }%
    }
    \end{center}

    We further test this result in multi-group societies: we configure our model with 4 groups $G_1, G_2, G_3, G_4$, with every group prejudiced against every other group, having agents distributed across them in the ratio $1:2:3:4$. Fig \ref{fig:4.1.4} shows the average \textit{prosperity} of each of these groups, and we observe that the \textit{prosperity} increases with an increase in group size. Hence our model and the derived result R3 is able to extend to known ideas of majorities in a society dominating over minorities, in the case of those majorities being prejudiced against the minorities.
    
\subsection{Presence of Renegades}
\label{subsec:3.3}

    \begin{figure}[h]
      \centering
      \begin{minipage}[t]{.45\textwidth}
        \centering
        \includegraphics[width=0.85\textwidth]{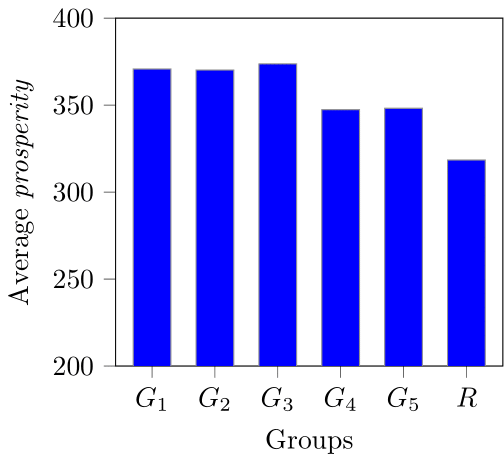}
        \captionof{figure}{Average \textit{Prosperity} levels in a multi-group society with groups containing \textit{renegades}. Here $G_4$ and $G_5$ have \textit{renegades}.}
        \label{fig:4.2.2}
      \end{minipage}\qquad
      \begin{minipage}[t]{.45\textwidth}
        \centering
        \includegraphics[width=0.85\textwidth]{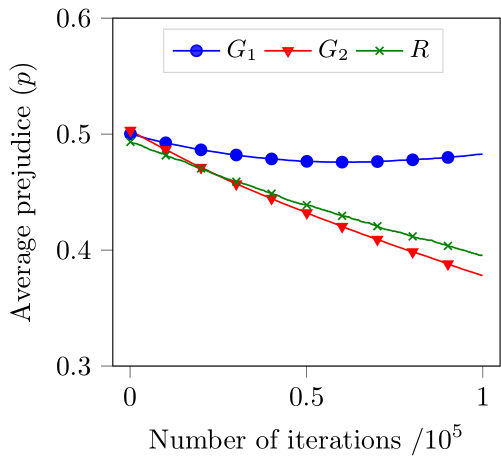}
        \captionof{figure}{Average prejudice levels in a two-group society with \textit{renegades}}
        \label{fig:4.2.1}
      \end{minipage}
    \end{figure}

    In societies, alongside prejudice, there can also exist an inherent and collective perception of some groups being superior to others. Theories suggest that in the perceived inferior groups in such societies, there may exist certain individuals who exercise out-group favoritism more than in-group favoritism~\cite{dasgupta_implicit_2004}. These individuals have been termed as \textit{renegades}~\cite{tajfel1974} and their impact on societies has not been addressed in existing approaches to modeling prejudice. Questions surrounding such agents remain largely unaddressed.
    
    In our framework, we model such agents as modified prejudiced agents with their prejudice being directed towards their own in-group rather than any other out-group. We conduct experiments to analyze the effect of the presence of renegade agents on the overall performance of the group they belong to, and also study how their prejudice levels and those of agents around them evolve over time.

    We run initial experiments on societies having two groups, $G_1$ and $G_2$, with each group being prejudiced against the other. While all agents in $G_1$ are prejudiced against $G_2$, some fraction of agents in $G_2$ are initialized as \textit{renegades}. When creating $G_2$ with only 10\% \textit{renegades}, we observe a 6.5\% decrease in the overall \textit{prosperity} of $G_2$ relative to $G_1$, and this number increases to 12.7\% when the population of renegades increases to 20\%. 
    
    Experimenting in the multi-group case, we initialize the model with 5 groups $G_1, ..., G_5$, with 10\% of the prejudiced agents in $G_4$ and $G_5$ being \textit{renegades}. Fig \ref{fig:4.2.2} shows the average payoffs of only the prejudiced agents in each group, along with that of all the renegades $(R)$ in our model. We observe that while the renegades themselves have the poorest performance of all agents, their presence even degrades the performance of other prejudiced agents in $G_4$ and $G_5$.
    
    We also track how the average prejudice values change over time in the two group experiment when 10\% of agents in $G_2$ are renegades. As shown in Fig \ref{fig:4.2.1}, while there is no significant evolution in the prejudice levels of $G_1$, agents in $G_2$ appear to have a reducing prejudice level over time, and this trend mirrors the prejudice level of the renegades themselves. These findings bring us to the following result:
    
    \begin{center}
    \fbox{%
    \parbox{0.8\columnwidth}{%
        \textbf{R4.} \emph{Renegades experience lower prosperity levels in society and they also bring about a reduction in prosperity for their fellow non-renegade group members. 
        Moreover, their existence implies conflicting prejudice orientations within a group, resulting in a decrease of both out-group prejudice in prejudiced agents and in-group prejudice in renegades.}
    }%
    }
    \end{center}


\subsection{Impact of Prejudice Levels Across \textit{Societies}}
\label{subsec:3.4}

    \begin{figure}[h]
      \centering
      \begin{minipage}[t]{.45\textwidth}
        \centering
        \includegraphics[width=0.85\textwidth]{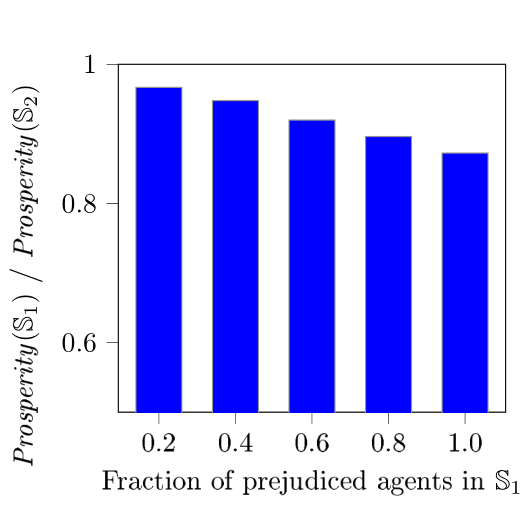}
        \captionof{figure}{\textit{Prosperity} ratio of prejudiced society $\mathbb{S}_1$ with unprejudiced society $\mathbb{S}_2$ for different concentrations of prejudiced agents in $\mathbb{S}_1$}
        \label{fig:4.3.1}
      \end{minipage}\qquad
      \begin{minipage}[t]{.45\textwidth}
        \centering
        \vspace{-13.7em}
        \include{tables/table_4}
        \vspace{4.85em}
        \captionof{table}{Average \textit{prosperity} values in societies with different number of groups containing prejudiced agents}
        \label{table:tab2}
      \end{minipage}
    \end{figure}

    Similar to Section \ref{subsec:3.1}, we analyze the effect of the level of prejudice. But instead of comparing different groups within a society, 
    we test this impact across different \textit{societies}.

    We run experiments comparing two societies, $\mathbb{S}_1$ and $\mathbb{S}_2$, each having 1000 agents. $\mathbb{S}_2$ is initialized with all agents as unprejudiced, whereas $\mathbb{S}_1$ is initialized with 5 different fractions of agents as prejudiced across 5 experiments: 0.2, 0.4, 0.6, 0.8, 1.0. Fig \ref{fig:4.3.1} plots the ratio of the average \textit{prosperity} of $\mathbb{S}_2$ to that of $\mathbb{S}_1$ for all 5 of these experiments. We observe that the payoff decreases as the percentage of prejudiced agents in $\mathbb{S}_1$ increases. This leads us to our final result: 
    
    \begin{center}
        \fbox{%
        \parbox{0.8\columnwidth}{%
            \textbf{R5.} \emph{As the number of prejudiced agents in a society increases, the average prosperity of that society as whole decreases.}
        }%
        }
    \end{center}
    
    We test this result for societies having more than two groups: We initialize a society with 5 groups, $G_1, G_2, G_3, G_4, G_5$, where each group has 200 agents.

    We start with the case where all groups are unprejudiced, and then run the model 5 more times, with one more group becoming prejudiced in each subsequent run.
    Table \ref{table:tab2} lists the average payoff value achieved by agents in each of these societies.
    We observe approximately a 20\% drop in \textit{prosperity} between a society with no prejudiced agents, and a society with all agents as prejudiced. 
    These findings reinforce our result R5, and also find credibility due to the universal understanding of prejudice being a negative entity in society.
    

\section{Discussions}

Experimental studies on prejudice, its origin, and its impact on people and groups exist dating back as far as 1970 \cite{tajfel1970}. Tajfel’s experiments, which were conducted on schoolboys aged 14 and 15, involved dividing the subjects into two groups based on flimsy and unimportant criteria, and then asking each subject to reward (in the form of money) some randomly chosen members that were explicitly labeled as belonging either to `their group' or to the `other group'.  The results showed that even though rationally, the best choices for individual subjects were to maximize the overall joint profit of all the boys combined, they instead made choices that favored the profits of their in-group over the out-group. The experiments noted that the subjects continued to make choices that were driven by an objective of maximizing the disparity between the profits of the groups, even if this disparity came at the cost of their own group getting lower profits than the theoretical maximum. Our results \textbf{R1} and \textbf{R5} corroborate the findings of these experiments, and we further demonstrate that this behavior exists in prejudiced societies even when the number of groups exceeds two. There also exist studies that do not regard prejudice as an abstract intergroup entity, but instead measure the impact of prejudicial discrimination, particularly in the context of interracial interactions \cite{crosby1980, Dovidio2002}. However, the implicit attitude of the prejudiced individuals to maximize the prosperity of their in-group remains a common denominator across all these studies. The fact that in-group favoritism (as demonstrated in result \textbf{R2}) and an objective of maximizing the in-group profits is an emergent behavior of the agents in our model speaks to the efficacy of the modeling methodology deployed by us for this task.

Studies with a similar experimental setup to Tajfel’s original work were published in 2001, with the objective being to study, amongst other things, the in-group satisfaction and discriminatory attitudes in individuals when there existed a significant imbalance in the sizes of the groups \cite{Leonardelli2001}. This study reached a similar conclusion to Tajfel’s work, as the subjects continued to make discriminatory choices to benefit their in-group over the out-group, even when they were members of the minority group. This result is also in line with our findings, as we show that prejudiced minorities achieve higher average prosperity levels when paired up with a non-prejudiced majority out-group (Fig. \ref{fig:4.1.3}). \cite{Leonardelli2001}, as well as other works, however, make no measurements to study a potential correlation between a group’s size and its performance. This is not necessarily surprising, since in most cases it is very difficult and sometimes even highly impractical to design and carry out experiments that test in isolation every such targeted hypothesis. This is where computational modeling, like the one presented in this work, is able to stand out over traditional experimentation, as it becomes possible to tailor experiments that test such hypotheses. We test the reasonably well-accepted wisdom of majorities being more socially prosperous than minorities through the experiments presented in Section \ref{subsec:3.2}. As shown through result \textbf{R3}, we find that the relative size of a group does play a role in its average prosperity levels in both two-group and multi-group societies. However, this remains true only when the groups in question are prejudiced and possess discriminatory attitudes, thus further highlighting the importance of isolating and identifying prejudice when studying any intergroup social phenomenon.

Although already well theorized by then \cite{Tajfel1971}, renegade agents, or agents that have a prejudicial attitude towards members of their in-group rather than their out-group, were first experimentally observed in 1980 when some individuals in a study by \cite{crosby1980} made decisions in certain social situations that benefited their racial out-group rather than their in-group. Existing studies and experiments on individuals possessing attitudes of out-group favoritism and self-stereotyping indicate a causal relationship between such attitudes and a negative impact on intellectual performance \cite{spicer}. These agents have also been modeled computationally in the past in the context of studying the role they play in bringing about polarization in societies \cite{FLACHE2018}. Experiments with these agent types have shown that even in low concentration (20 percent), they can significantly influence the members of their in-group to have a more positive outlook toward members of the out-group and reduce the overall out-group hatred. Similar results on the impact of renegades on opinions are obtained from our model as well, as shown by the decrease in prejudice levels of the non-renegade members belonging to the renegades’ in-group (Fig \ref{fig:4.2.1}). Existing models, however, exclusively study the role these agent types play in changing the opinions of other agents, and fail to capture and predict their potential impact on the performance (or prosperity) of their in-group in social interactions. \cite{dasgupta_implicit_2004} hypothesizes that the existence of in-group prejudice may negatively impact the performance of both the self-stereotyping individuals and their in-group. Our model allows us to design experiments that can isolate and measure the impact of renegade agent types which lets us not only confirm, but also further refine existing hypotheses. In our simulations, these agent types experience the lowest prosperity levels in society, and can lead to decreased prosperity for the rest of their in-group compared to the non-self-stereotyping out-group(s) in both the two-group and multi-group settings (Section \ref{subsec:3.3}).

The group division methodology deployed by \cite{tajfel1970} for his experimental studies has since been widely accepted in the research community and is referred to as the minimal group paradigm \cite{Tajfel1971}. What these experiments highlight is that prejudicial attitudes can often emerge implicitly as a side-effect of group division itself, and may not necessarily stem only from deep-rooted issues such as race or gender. In experimental studies of any intergroup phenomenon, it is nearly impossible to isolate the effect of the phenomenon under study from the implicit prejudices that may arise between the groups. Therefore, in computational modeling, in order to accurately emulate real-world outcomes, it becomes critical to account for this implicit prejudice. This is where the model presented by us in this work can be deployed by researchers in conjunction with their own for the phenomenon under study, and thus either incorporate or remove the impact of implicit prejudices.

\section{Conclusion}

    Prejudice is crucial in understanding multi-group dynamics as it provides insights into the kind of inter-group relations prevalent in a social setting. Our framework introduces an agent type, the prejudiced agent, to analyze the relationship between prejudice and prosperity by creating societies containing groups and factions.  
    
    Previous studies on modeling out-group prejudice focus more on the evolution and propagation of prejudice in society~\cite{Whitaker2018}. In this work, we present a model that not only tracks the changes in prejudice levels of individuals but also the effect it can have on an agent's individual and group prosperity. We show that even when modeling prejudice as out group-suppression, its existence results in the implicit emergence of in-group promotion, making these concepts impossible to separate from each other. Our results suggest that while prejudiced agents themselves accumulate relatively higher prosperity levels, their presence in societies reduces the prosperity level as a whole. We also show that prejudiced groups in a society maintain higher levels of prosperity even when they are a minority in the society.
    
    The model proposed in this work can serve as a strong foundation to conduct meaningful simulation based studies on prevalent social issues. By appropriate instantiation of prejudiced agents, factions and groups, one can analyze the effects of prejudice in interactions between people of different races, nationalities, etc. Much research has already been conducted on studying and preventing social issues such as racism which have continued to plague our society even in this modern era. However, when it comes to the testing and applicability of such theoretical solutions, there is a clear lacuna due to the difficulty inherent in conducting social experiments. We believe that our model can form a basis by which these theories can be easily simulated and tested.
    
    Existing experiments on intergroup phenomena follow Tajfel’s minimal group paradigm, which assumes group alignment to be a one-dimensional trait. This means that existing studies do not account for two agents that may be part of the same group for an aspect (e.g. politics) and at the same time belong to different groups in other respects (e.g. race). When such real-world studies are conducted, this model can be correspondingly enhanced.
    
    This model may be enriched with economic aspects for the purpose of studying other social phenomena of interest, such as the distribution of wealth in a prejudiced society. An agent in a prejudiced environment would undergo preferential interactions in the sense that it would act upon its prejudice to decide its partners for economic transactions. Such an economic model could then be used to conduct extensive experiments to learn about the effects of prejudice on the social status of a person by taking the wealth accumulated by an agent as a proxy for its social status.

\bibliographystyle{unsrt}  
\bibliography{references}

\end{document}

%% file: tables/cpd.tex
\begin{table}[h]
    \centering
    \begin{tabular}{llcc}
                                                          &                        & \multicolumn{2}{c}{Agent $\mathcal{A}{}_1$}                         \\
                                                          &                        & 1                             & 0                             \\ \cline{3-4} 
    \multicolumn{1}{c}{\multirow{2}{*}{Agent $\mathcal{A}{}_0$}} & \multicolumn{1}{c|}{1} & \multicolumn{1}{c|}{$(C, C)$} & \multicolumn{1}{c|}{$(S, T)$} \\ \cline{3-4} 
    \multicolumn{1}{c}{}                                   & \multicolumn{1}{c|}{0} & \multicolumn{1}{c|}{$(T, S)$} & \multicolumn{1}{c|}{$(D, D)$} \\ \cline{3-4} 
    \end{tabular}
    \end{table}

%% file: tables/table_1.tex
\begin{table}[ht]
\centering 
\begin{tabular}{c l} 
\hline 
\textbf{Variable} &\textbf{Description} \\ [0.5ex]
\hline 
$\omega$ &Memory of an agent \\
\textit{f} &Faction Alignment of an agent \\
\textit{$p_{i}$} &Prejudice value of an agent against members of Group \textit{i} \\
\textit{$Prej_F$} &Faction \textit{$\mathcal{F}$'s} prejudice against members of a Group\\
$\eta_\mathcal{A}{}_0(\mathcal{A}{}_1, t)$ &Opinion of $\mathcal{A}{}_0$ about $\mathcal{A}{}_1$ at interaction \textit{t} \\ 
$\mathcal{O}_\mathcal{F}(\mathcal{A}{}_1, t)$ &Faction $\mathcal{F}$'s Aggregated Opinion of $\mathcal{A}{}_1$ at interaction $t$\\
$\Psi_{\mathcal{A}{}_0}(\mathcal{A}{}_1,t)$ &Prejudiced Opinion of $\mathcal{A}{}_0$ about $\mathcal{A}{}_1$ at interaction \textit{t}\\
$\mathcal{P}_\mathcal{F}(\mathcal{A}{}_1, t)$ &Faction $\mathcal{F}$'s Prejudiced Aggregated Opinion of $\mathcal{A}{}_1$ at interaction $t$\\
$\mathcal{C}_{\mathcal{A}{}_0}(\mathcal{A}{}_1, t)$ &The Cooperation Level for $\mathcal{A}{}_0$ at interaction \textit{t} with $\mathcal{A}{}_1$\\[1ex] 
\hline 
\end{tabular}
\vspace{0.5em}
\caption{Notation used} 
\label{table:variables}
\end{table}

%% file: tables/table_2.tex
\begin{table}[ht]
\centering
    \begin{tabular}{lll}
    \hline
    \textbf{Parameter}                 & \textbf{Description}                                                                                                   & \textbf{Value}                      \\ \hline
    $n$      & Number of agents in the model                                                                                 & 1000                       \\
    $n_f$ & Number of agents in each faction                                                                              & 20                         \\
    $\omega$ & Memory size of each agent                                                                                 & $10$ \\
    $\alpha_1$           & \begin{tabular}[c]{@{}l@{}}Threshold value for payoff in order\\ to increase the prejudice level\end{tabular} & $2$                     \\
    $\alpha_2$           & \begin{tabular}[c]{@{}l@{}}Threshold value for payoff in order\\ to decrease the prejudice level\end{tabular} & $1.5$                     \\
    $\theta$              & Prejudice update step size                                                                                      & $0.005$                    \\
    $\beta_1$           & \begin{tabular}[c]{@{}l@{}}Threshold value for increasing \\the faction alignment\end{tabular} & $0.01 \leq \beta_1 < 0.08$                     \\
    $\beta_2$           & \begin{tabular}[c]{@{}l@{}}Threshold value for decreasing \\the faction alignment\end{tabular} & $0.08 \leq \beta_2 \leq 0.2$                     \\
    $\tau$              & Faction alignment update step size                                                                                      & $0.005$                    \\
\hline
    \end{tabular}
    \vspace{0.5em}
    \caption{Model Parameters} 
    \label{table:model}
    \end{table}

%% file: tables/table_3.tex
\begin{tabular}{ll}
\hline
Group    & Groups Prejudiced Against \\ \hline
$G_1$    & -                         \\
$G_2$    & -                         \\
$G_3$    & -                         \\
$G_4$    & -                         \\
$G_5$    & -                         \\
$G_6$    & $G_1$                     \\
$G_7$    & $G_1, G_2$                \\
$G_8$    & $G_1,G_2,G_3$             \\
$G_9$    & $G_1, G_2, G_3,G_4$       \\
$G_{10}$ & $G_1, G_2, G_3, G_4, G_5$
\end{tabular}

%% file: tables/table_4.tex
\begin{tabular}{ll}
\hline
Groups prejudiced & \begin{tabular}[c]{@{}l@{}}\textit{Prosperity}\end{tabular} \\ \hline
-                 & 448                                                      \\
$G_1$              & 429                                                      \\
$G_1, G_2$          & 411                                                    \\
$G_1, G_2, G_3$      & 393                                                      \\
$G_1,G_2, G_3, G_4$   & 375                                                      \\
$G_1,G_2,G_3,G_4, G_5$ & 359                                                      \\ \hline
\end{tabular}